\documentclass[12pt]{article} 

\usepackage[hidelinks]{hyperref}
\hypersetup{allcolors=blueish,colorlinks=true}

\usepackage{siunitx}
\usepackage{threeparttable}

\usepackage{scicite}

\usepackage{times}

\usepackage{amsmath}
\usepackage{amsfonts}
\usepackage{amssymb}
\usepackage{graphicx}

\usepackage{booktabs}

\newenvironment{sciabstract}{%
\begin{quote} \bf}
{\end{quote}}

\usepackage[labelfont=bf,figurename=Fig.,labelsep=period,font=footnotesize]{caption}

\title{\Large\bf Larger cities, more commuters, more crime? The role of inter-city commuting in the scaling of urban crime}

 \author
 {
Simon Puttock,${}^{1}$ 
 Umberto Barros,${}^{2}$ \\
 Diego Pinheiro,${}^{2}$ 
 and Marcos Oliveira${}^{1,3\dagger}$\vspace{.2em}\\
 \small{${}^{1}$University of Exeter, Exeter, UK}\\
 \small{${}^{2}$Universidade de Pernambuco, Recife, Brazil}\\
 \small{${}^{3}$Vrije Universiteit Amsterdam, Amsterdam, Netherlands}\vspace{.1em}\\
 \footnotesize{
 $^\dagger$\href{mailto:m.a.oliveira@vu.nl}{m.a.oliveira@vu.nl}
 }\vspace{-2em}
 }

\date{}

\usepackage{changepage}

\usepackage{xcolor}
\usepackage{soul}
\definecolor{reddish}{HTML}{FBB4AE}
\definecolor{blueish}{HTML}{1F54A9}
\definecolor{magentish}{HTML}{FF00AA}
\definecolor{greenish}{HTML}{a1d99b}

\usepackage[hidelinks]{hyperref} 

\usepackage{setspace}

\usepackage{lineno}

\usepackage{lipsum}

\begin{document} 




\maketitle 




\begin{sciabstract}
Abstract.

\textnormal{%
%
Cities attract a daily influx of non-resident commuters, reflecting their roles within wider urban networks---not as isolated places. However, it remains unclear how this inter\-connectivity shapes the way crime scales with population, given that larger cities tend to receive more commuters and experience more crime. In this work, we investigate how inter-city commuting relates to the population--crime \mbox{relationship}. We find that larger cities receive proportionately more commuters, which in turn is associated with higher levels of burglary, drug possession, robbery, shoplifting, and theft. For example, each 1\% increase in inbound commuters corresponds to a 0.32\% rise in theft and 0.20\% rise in burglary, holding population size constant. We demonstrate that models incorporating both population size and commuter inflows explain variation in these offenses better than population-only models. Our findings underscore the importance of considering how cities are connected---not just their population size---in disentangling the population--crime relationship.
}
\end{sciabstract}
\baselineskip1.8em 

\section*{Introduction}
Crime tends to occur more frequently in more populated places---a regularity that underpins criminology theories and the widespread use of per capita crime rates to compare cities\cite{Chamlin2004,Rotolo2006,oliveira2021more}. However, cities are not closed systems: they draw large daily inflows of non-residents, such as commuters and tourists, who contribute to urban activity and raise the number of people in circulation\cite{Gibbs1976,stafford_crime_1980,StultsHasbrouck}. Such flows reflect the role of cities as nodes within broader urban networks, where hubs attract and concentrate economic, social, and cultural activity\cite{neal_connected_2013,liang_intercity_2024}. Yet it remains unclear how this inter-city connectivity factors into the scaling of crime with population size.

The relationship between crime and population is a long-standing empirical observation, with criminology theories---structural, social control, and subcultural---linking variations in population size to shifts in social dynamics that influence crime\cite{Chamlin2004,Rotolo2006,oliveira2021more}. Specifically, structural theories suggest that more people mean more interactions and, with them, more opportunities for conflict\cite{Mayhew1976,blau1977inequality}. From a different angle, social control theories argue that dense, heterogeneous populations weaken informal mechanisms that inhibit deviance\cite{wirth1938urbanism,Groff2015}. In contrast, subcultural theories propose that larger populations foster the emergence of alternative social norms that can support criminal behavior\cite{fischer1975toward,Fischer1995}. Regardless of theoretical framing, the population--crime relationship has been consistently observed, and recent studies have shown that the link is not always proportional; certain offenses, such as theft and violent crime, increase faster than population, exhibiting increasing returns to scale\cite{Bettencourt2007,oliveira2021more}. This scaling may reflect not only population size, but also how cities are connected to one another.

Rather than isolated entities, cities are part of networks of exchange---of goods, people, and ideas---and attract outsiders through economic and social opportunities. Early works have recognized that ``singular cities" concentrate inflows from their surroundings, drawing in individuals who may face increased risks of offending or victimization\cite{Gibbs1976,stafford_crime_1980,farley_ecological_1981,farley_suburbanization_1987}. This inter-city dependence means that city-level crime analyses ignoring meso- or micro-level relationships risk producing bias\cite{oliveira2017scaling,hipp_cities_2017,hipp_accounting_2021}. Likewise, daily inflows in cities create a discrepancy between residential and ambient populations, and neglecting them can distort crime rates\cite{StultsHasbrouck}, a mismatch also evident at finer scales where crime scales with floating populations\cite{caminha_human_2017}.
To address this distortion, previous works have suggested estimating ambient population and accounting for the number of targets\cite{Boggs1965,Andresen2006,Andresen2011}. Yet, despite these methodological advances, prior studies have failed to understand whether and how inter-city connectivity itself relates to the relationship between population size and crime.

Insights from other domains, however, suggest that city connectivity can significantly shape how urban indicators scale with population. In the economy, both national commuting flows\cite{alves_commuting_2021} and global connectivity\cite{yang_inter-city_2024} help explain why some cities exceed GDP predictions based on city size alone. Similar effects appear in innovation, where inter-city firm connectivity improves population-based models of patenting activity\cite{liang_intercity_2024}. Likewise, in public health, disease incidence across cities scales not only with population but also with the strength of commuting ties\cite{loureiro_impact_2025}. Such findings from urban science demonstrate that city connectivity plays a fundamental role in how urban indicators scale---yet this perspective remains unexplored in understanding how crime scales with population.
 
Here, we investigate how inter-city commuting relates to the population--crime {relationship}. We use data on 8 million workers in England and Wales to construct a commuting network and assess how crime scales with population and inbound commuters. We find that inbound commuting grows proportionally with city size, and that models incorporating both population and commuter inflows provide a more parsimonious and better-fitting explanation of crime levels than models based on population alone for burglary, drug possession, robbery, shoplifting, and theft. In contrast, drug trafficking and homicide show no evidence of association with inter-city connectivity. For the offenses associated to commuting, we find that crime levels scale superlinearly with the combined effect of population and commuting---but once commuter inflows are accounted for, the relationship between crime and population becomes approximately linear. Our work demonstrates that inter-city connectivity is a crucial factor in crime scaling, revealing that disentangling the population--crime relationship requires considering how cities are connected to one another.

\section*{Results}
To investigate the relationship between commuting patterns, population size, and crime, we use data from England and Wales, combining commuting data from the UK Census with police-recorded crime data. Both data sets cover the same period and are aggregated at the level of Community Safety Partnerships (see Methods for more details on data sources and processing).

 
\begin{figure}[b!]
\centering
\includegraphics[width=6.25in,height=4.5in]{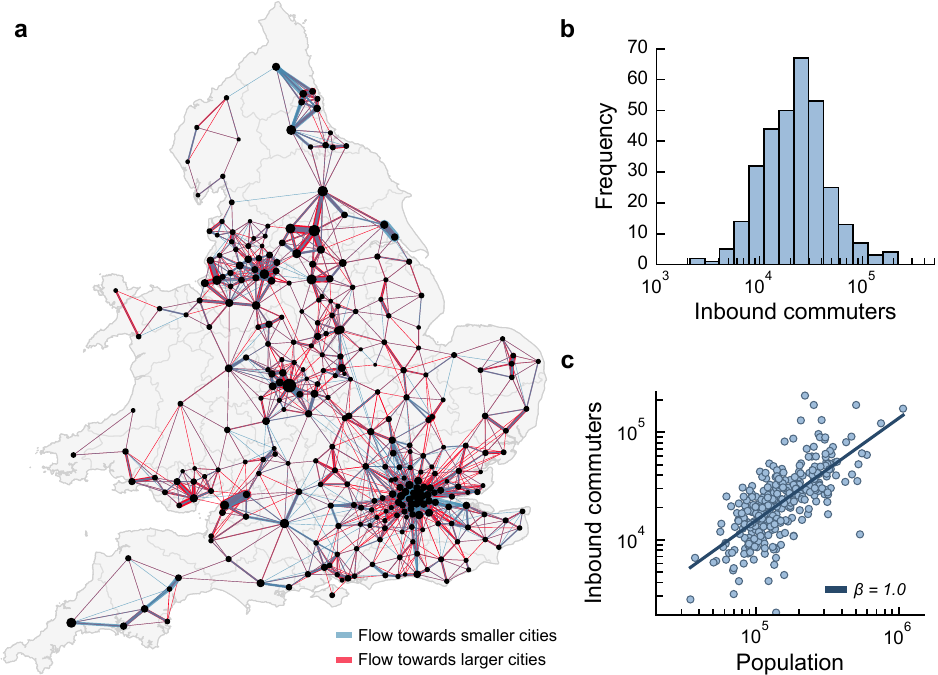}
  \caption{\textbf{The inter-city commuting network.} 
  \textbf{(a)} We construct a commuting network where nodes represent cities, and directed edges indicate the number of individuals traveling between them for work. In the plot, edge color shows direction, and thickness reflects the number of commuters (only flows above 500 are shown). Most commuting flows are spatially constrained, occurring between nearby areas, but some cities attract disproportionately large inflows, 
  \textbf{(b)}~leading to a distribution of inbound commuters with broad variance. While most cities receive a moderate number of workers, a few attract substantially more, forming a long right tail.  
  \textbf{(c)}~This attraction is tied to city size: more populated cities receive more commuters, and this relationship is approximately linear.
      }
\label{fig1}
\end{figure}

\subsection*{Larger cities, more inbound commuters}
Before examining how commuting patterns affect crime, we first analyze the distribution of inbound commuters---workers who regularly travel from their place of residence to another area for work (i.e., crossing administrative boundaries).  To do this, we construct a commuting network where nodes represent cities and directed edges capture the number of individuals traveling from one place to another for work (Fig.~\ref{fig1}a). This network shows that commuting is largely spatially constrained: most flows occur between nearby areas. It also reveals a broad variance in the number of inbound commuters, which approximately follows a log-normal distribution with a median of 22,997 (Fig.~\ref{fig1}b). Most places attract a moderate number of commuters, but a few receive substantially more, forming a long right tail in the distribution. For example, Leeds receives around 121,278 commuters---nearly one for every six residents, given a population of approximately 751,485. To better understand this broad variance, we examine how the number of inbound commuters scales with population size.

We find that more populated cities receive more inbound commuters, and this growth is proportional to their population size rather than outpacing it. To quantify this relationship, we estimate the scaling exponent $\beta$ from $I \sim N^{\beta},$ where $I$ is the number of inbound commuters and $N$ is the population size (see Methods). Our results show that inbound commuting scales approximately linearly with population (i.e., $\beta \approx 1$), indicating that the number of inbound commuters increases proportionally with population (Fig.~\ref{fig1}c). This growth contrasts with the superlinearity observed in other urban indicators (e.g., crime, income) and may reflect constraints at the individual level---such as time, cost, or effort---that prevent commuting distance from outpacing population growth\cite{louf_how_2014}. In any case, this scaling represents a steady inflow of people and suggests that more populated places are more attractive to outsiders; we next examine how this relates to the crime--population relationship.


\subsection*{Larger cities, more commuters, more crime}

\begin{figure}[b!]
\centering
\includegraphics[width=6.25in,height=4.25in]{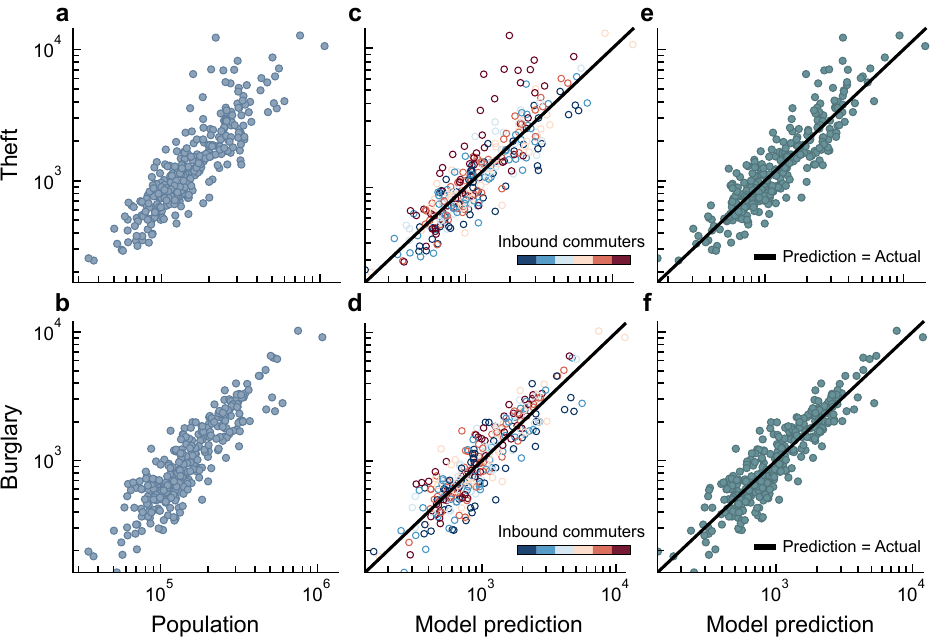}
  \caption{\textbf{The role of inter-city commuting on the crime--population relationship.} 
  \textbf{(a, b)} Theft and burglary both scale superlinearly with population: as city size increases, these crimes rise faster than proportionally.
  Some of the variation in crime not captured by population alone can be explained by \textbf{(c, d)} incorporating commuter inflows. In the plots, dots are the prediction of the population-only model, whereas their color represents the percentile of inbound commuters (given population size).
  \textbf{(e, f)}~We find that a population-and-commuter model yields more parsimonious and better-fitting explanation than the population-only model.
  }
\label{fig2}
\end{figure}

To set a baseline for understanding how commuter flows relate to crime, we first establish how crime scales with population size, replicating the population--crime relationship reported in previous studies. We estimate the scaling exponent $\beta$ from $C \sim N^{\beta}$, where $C$ is the number of crimes and $N$ is the population size, focusing on theft and burglary first, before turning to other crime types (see Methods). Our estimates align with prior findings for the UK\cite{oliveira2021more}. We find that both crimes scale superlinearly with population, with $\beta = 1.22$ (95\% CI: 1.14--1.29) for theft and $\beta = 1.25$ (95\% CI: 1.18--1.32) for burglary (Fig.~\ref{fig2}ab). This superlinearity implies that a 1\% increase in population is associated with more than a 1\% increase in crime---a 1\% population growth corresponds to a 1.22\% increase in theft and 1.25\% in burglary.


Larger cities have more crime---and more commuters. Next, we disentangle how crime scales with both population and inbound commuters (Fig.~\ref{fig2}cd). To do so, we extend our model to include both resident population and inbound commuters. We model crime as a function of population and commuter inflows, $C \sim N^{\gamma} I^{\alpha}$, where $N$ is the population size and $I$ is the number of inbound commuters. This formulation follows the Cobb–Douglas production function, commonly used to capture the joint contribution of multiple factors\cite{cobb_theory_1928,varian_intermediate_2014}, and builds on prior applications of production-function models to urban output\cite{lobo_urban_2013,ribeiro_effects_2019}.
We estimate the exponents $\gamma$ and $\alpha$ from the data and compare this model against the population-only version using the Bayesian Information Criterion (BIC). We find that the population-and-commuter model provides a more parsimonious and accurate fit. For theft, we find $\gamma = 0.92$ (95\% CI: [0.83, 1.01]) and $\alpha = 0.32$ (95\% CI: [0.25, 0.39]); for burglary, $\gamma = 1.06$ (95\% CI: [0.97, 1.15]) and $\alpha = 0.20$ (95\% CI: [0.14, 0.27]) (see Fig.~\ref{fig2}ef). These estimates, along with the better fit of the model relative to the population-only version, suggest that population and inter-city commuting jointly contribute to the observed scaling of urban crime. 

In both theft and burglary, crime scales superlinearly with the combined effects of population and inbound commuters (i.e., $\gamma + \alpha > 1$). However, once we adjust for inbound commuting, the relationship between crime and population becomes approximately linear, as the confidence intervals of $\gamma$ include $1$. When isolating the effect of inbound commuting, our results suggest that theft is slightly more sensitive to commuter inflows than burglary. Our model indicates that a 1\% increase in inbound commuters is associated with a 0.32\% rise in theft and a 0.20\% rise in burglary, holding population constant.
We also compared the population-and-commuter model to a model specification that includes interaction effects but found that the simpler model provided a more parsimonious and better fit (see Methods).

\begin{figure}[h!]
\centering
\includegraphics[width=3.4in,height=3.5in]{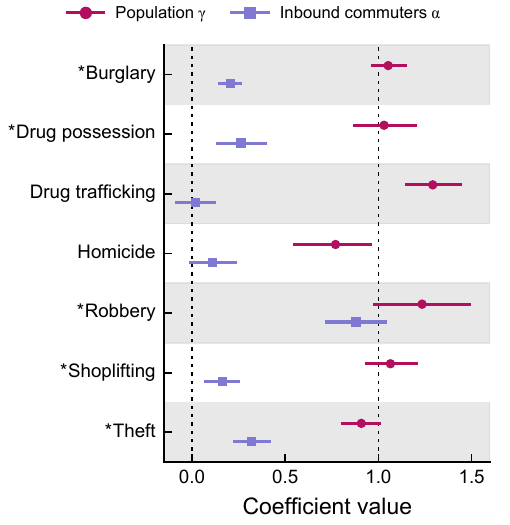}
  \caption{\textbf{
  The role of inter-city commuting across crime types.%
  } 
  We estimate population and inbound commuter coefficients for each offense using the population-and-commuter model.
  This model explains variation 
  in burglary, drug possession, robbery, shoplifting, and theft; in contrast, drug trafficking and homicide show no evidence of association with inbound commuting. The asterisks indicate where the population-and-commuter model offers a better fit than the population-only specification, whereas the bars denote 95\% confidence intervals.
  }
\label{fig3}
\end{figure}

Next, we extend our analysis to other crime types---drug possession, drug trafficking, homicide, robbery, and shoplifting---to examine how the influence of commuting varies across offenses. Our results reveal that the population-and-commuter model outperforms the population-only model for drug possession, robbery, and shoplifting; in contrast, the population-only model provides a better fit for drug trafficking and homicide (see Fig.~\ref{fig3}). We find  $\gamma = 1.05$ (95\% CI [0.91, 1.20]) and $\alpha = 0.25$ (95\% CI [0.15, 0.36]) for drug possession; $\gamma = 1.25$ (95\% CI [0.99, 1.50]) and $\alpha = 0.89$ (95\% CI [0.70, 1.08]) for robbery; and $\gamma = 1.08$ (95\% CI [0.94, 1.22]) and $\alpha = 0.16$ (95\% CI [0.05, 0.26]) for shoplifting (see Methods). Similar to theft and burglary, these offenses scale superlinearly with population but become approximately linear once commuting is included. Notably, drug trafficking and possession diverge: though both scale superlinearly with population ($\beta \approx 1.3$), only possession is associated with inbound commuting, highlighting their distinct underlying mechanisms, 
otherwise hidden by population-only scaling. 

\section*{Discussion}

Cities are not isolated systems; they are embedded in broader urban networks that shape how people live, work, and interact. Though prior works have recognized that this interconnectivity affects crime, they have failed to quantify how it relates to the population--crime relationship. In this work, we analyze inter-city commuting data to investigate how connectivity between cities influences the scaling of crime with population size.

First, we used census data from 8 million workers in England and Wales to construct an inter-city commuting network; then, we evaluated how the number of inbound commuters is associated with the relationship between population and crime, specifically theft and burglary. By comparing different model specifications, we examine whether incorporating commuter flows better explains variations in crime across cities than population alone.

We find that models including both population size and inter-city commuting explain crime variations more accurately and parsimoniously than population-only models in the case of burglary, drug possession, robbery, shoplifting, and theft. We show that, though these offenses scale superlinearly with population, the relationship becomes linear once we adjust for inbound commuters. When we isolate the effect of commuting, we find that higher inbound flows are associated with increased crime. 

In contrast, we find no evidence that inbound commuting is associated with homicide or drug trafficking. In particular, it is noteworthy that drug trafficking and drug possession yield distinct results. Though both scale superlinearly with population, only possession is associated with inbound commuting, bringing out their distinct underlying mechanisms. This distinction would remain unobserved in traditional population-only scaling analyses, showing that superlinearity can arise with or without inter-city connectivity and demonstrating the need to account for connectivity when examining the population--crime relationship.

Our findings support early ecological perspectives that view cities as interdependent rather than isolated units, where dominant centers attract nonresidents and expand the pool of potential offenders and victims beyond the resident population\cite{Gibbs1976,stafford_crime_1980}. These studies argued that variation in city crime rates reflects ecological position and dominance within the metropolitan system rather than solely local social or economic conditions. Our results provide quantitative evidence for this view, suggesting that analyses of urban crime must account for both local conditions and inter-city dynamics. They also corroborate recent evidence that neglecting meso-level processes can bias city-level analyses\cite{hipp_accounting_2021}. Together, these findings reinforce that population-only models miss the broader urban context. By combining scaling analysis with commuting data, we explicitly incorporate inter-city connectivity and quantify how commuter inflows shape the scaling of crime with population size.


These results also have broader implications for the theories of urban scaling. Most mechanistic models describe superlinear growth as the outcome of processes occurring within cities, such as intensified social interactions and increasing functional diversity\cite{ribeiro_mathematical_2023}, with a few notable exceptions that incorporate inter-city processes\cite{leitao2016scaling}. Our results show that inter-city connectivity contributes to scaling patterns, corroborating recent evidence of spatial and network effects linking urban systems\cite{yang_inter-city_2024,altmann_spatial_2020}. The association between crime and commuter inflows indicates that processes shaping urban outcomes operate across, not only within, city boundaries, suggesting that mechanistic models of crime should account for inter-city connectivity.

In summary, our work advances our understanding of the population--crime relationship by disentangling its association with inter-city connectivity. Our findings call for more nuanced approaches that explicitly model cities not as isolated entities but as interconnected nodes within broader networks. Future research on crime modeling should therefore incorporate both local population dynamics and inter-city interactions to situate crime within broader urban systems.

\subsection*{Limitations}

Our work presents limitations related to city definition and data sets. 
First, we note that urban scaling analyses depend on how cities are defined, as definitions \mbox{vary---from} legal units (e.g., counties, municipalities) to data-driven delineations based on population density and economic interactions---and can yield divergent scaling exponents for the same indicator\cite{Arcaute2014,Cottineau2017,Louf2014}. In our analysis, we rely on data at the level of Community Safety Partnerships---local administrative units that coordinate crime prevention strategies and comprise police, local authorities, and other agencies. We refer to these units as cities or regions throughout the text. While they may align with legal city boundaries or functional definitions, this is not guaranteed, as local partnerships primarily shape them. Nonetheless, these units are highly relevant from both a policy and data perspective: many crime reduction strategies are designed, implemented, and assessed at this level. 
Future research should explore alternative definitions, such as demographic clustering to define communities\cite{white2014spatial} and network-based methods that link areas by crime incidence\cite{10.1371/journal.pone.0143638,oliveira2015criminal}, to test the sensitivity of findings to spatial boundaries.
Second, our analysis is constrained by the availability of commuting data, which comes from the 2011 United Kingdom census---the most recent official source. Although more recent data are unavailable, future work could explore novel sources, such as mobile phone records or social media, to capture more up-to-date commuting patterns\cite{Malleson2016,PachecoOM17}.

\section*{Methods}

\subsection*{Data}

This paper uses commuting data from the United Kingdom's 2011 Census ``Place of Residence by Place of Work, Local Authority" data set\cite{Commutedataset}, which records commuters flows between local authority districts in an adjacency matrix-style table. Each row and column represents residential and work locations, respectively, with the cell values indicating the number of commuters. 
A limitation of this data set is that local authority districts merge smaller towns and villages into single nodes, obscuring finer commuting patterns, particularly in the areas between major metropolitan hubs. Another issue pertains to the geography of the British Isles: Northern Ireland and Great Britain are separated by the Irish Sea, and commuting between them is minimal. As a result, we exclude Northern Ireland from this analysis.

To explore the relationship between crime and commuting, we use police-recorded crime data from 2011/2012 to match the census period. Recording practices vary across police forces and between countries in the UK. In England and Wales, crime data is attributed to Community Safety Partnerships (CSPs), local collaborations that aim to reduce crime and improve community safety. For our analysis, we use the ``Police Recorded Crime Community Safety Partnership Open Data Table, year ending March 2012'' data set, which provides detailed records of all recorded offences per CSP from April 2011 to March 2012~\cite{Crimedataset}.
Because this data set is more detailed and consistent than those available for other parts of the UK, we focus exclusively on England and Wales, excluding Scotland and Northern Ireland from our crime analysis. Additionally, since some CSPs encompass multiple local authorities---for example, North Yorkshire CSP includes Craven, Hambleton, Harrogate, Richmondshire, Ryedale, Scarborough, and Selby---we pre-processed the data to aggregate nodes from the local authority commuting data network to their corresponding CSPs used in the 2011 crime data set. To associate each LAD with its respective CSP, we use the ``LAD to Community Safety Partnership to PFA (December 2019)'' data set\cite{LADtoCSPdataset}, which provides the most consistent structural match with the 2011 Census commuting data. The ``UK Local Authorities (past and current)'' data set\cite{LADInformationDataset} was used to obtain LAD codes---essential for linking LADs to CSPs---and to retrieve geographic coordinates for constructing the England and Wales network. Finally, we used the ``2011 Census: Usual resident population, local authorities in England and Wales'' to obtain the estimated total population for each LAD\cite{PopulationDataset}, which we preprocessed by aggregating multiple LADs into their corresponding CSPs and summing their populations where CSPs spanned more than one LAD.
%

\subsection*{Regression}
To estimate how crime scales with population and inbound commuting, we use linear regression on log-transformed data. We begin with the standard urban scaling model $C \sim N^{\beta}$,
where $C$ is the number of crimes and $N$ is the resident population (Model 1). We linearize this using logarithms, obtaining
\begin{equation*}
    \log C = \beta \log N + \varepsilon,
\end{equation*}
and estimate the scaling exponent $\beta$ using ordinary least squares (OLS). 
To incorporate commuting, we extend this to a two-factor Cobb--Douglas form:
$C \sim N^{\gamma} I^{\alpha}$,
where $I$ denotes the number of inbound commuters (Model 2). This yields the linearized model
\begin{equation*}
\log C = \gamma \log N + \alpha \log I + \varepsilon,
\end{equation*}
with parameters again estimated via OLS.
To evaluate model performance, we compare the population-only and population-and-commuter models using the Bayesian Information Criterion (BIC), which balances model fit and complexity. A lower BIC indicates a more parsimonious and better-fitting model.
Additionally, we test a third model specification (Model 3) that includes an interaction term between population and commuting:
\begin{equation*}
    \log C = \gamma \log N + \alpha \log I + \delta (\log N \times \log I) + \varepsilon.
\end{equation*}
For burglary, Model 2 offers the best fit based on BIC and includes only statistically significant predictors (Table~\ref{tab:regression_models_burglary}). For theft, Model 3 slightly outperforms Model 2 in BIC but only marginally; given the minimal gain in explanatory power and added complexity, Model 2 remains the most robust and interpretable choice (Table~\ref{tab:regression_models_theft}). Supplementary Note 1 provides the tables for the remaining offenses.

We note that resident population and inbound commuting are correlated across cities, but diagnostics show that this does not pose a problem of multicollinearity in Model~2. The condition number of the design matrix is $158$ and the variance inflation factor (VIF) for the regressors is $1.96$, values that indicate moderate correlation but no regression instability and negligible variance inflation. As an additional robustness check, we re-estimated the model using ridge regression, with the penalty parameter $\lambda$ selected by leave-one-out cross-validation (see Supplementary Note 2), following the procedure of previous studies\cite{ribeiro_effects_2019,alves_commuting_2021,loureiro_impact_2025}.
We find that the ridge regression coefficients are very similar to the OLS estimates.
For example, for theft, we find with ridge regression $\gamma = 0.91$ (95\% CI [0.80, 1.01]) and $\alpha = 0.32$ (95\% CI [0.22, 0.42]), whereas OLS estimates are $\gamma = 0.92$ (95\% CI [0.83, 1.01]) and $\alpha = 0.32$ (95\% CI [0.25, 0.39]), and, for burglary, we find with ridge regression $\gamma = 1.05$ (95\% CI [0.96, 1.15]) and $\alpha = 0.21$ (95\% CI [0.14, 0.27]), whereas OLS estimates are $\gamma = 1.06$ (95\% CI [0.97, 1.15]) and $\alpha = 0.20$ (95\% CI [0.14, 0.27]). 
This similarity is also the case for all the other offenses and demonstrates that multicollinearity does not affect coefficient stability in our analysis, with ridge estimates being just slightly smaller than OLS, as expected due to penalization.

\begin{table}[h!]
\centering
\caption{Comparison of OLS Regression Models for Burglary.}
\label{tab:regression_models_burglary}
\hspace*{\fill}
\begin{tabular*}{0.88\textwidth}{l@{\extracolsep{\fill}}*{3}{S[table-format=-1.3]@{\,}l}}
\toprule
 & \multicolumn{2}{c}{{Model 1}} & \multicolumn{2}{c}{{Model 2}} & \multicolumn{2}{c}{{Model 3}} \\
\midrule
Intercept & -3.502\textsuperscript{***} & ($0.179$) & -3.384\textsuperscript{***} & ($0.171$) & 0.037 & ($1.974$) \\
$N$ & 1.255\textsuperscript{***} & ($0.035$) & 1.060\textsuperscript{***} & ($0.046$) & 0.397 & ($0.384$) \\
$I$ &  &  & 0.205\textsuperscript{***} & ($0.034$) & -0.571 & ($0.447$) \\
$N \times I$ &  &  &  &  & 0.150\textsuperscript{*} & ($0.086$) \\
\midrule
R\textsuperscript{2} & \multicolumn{2}{S[table-format=-1.3]}{0.805} & \multicolumn{2}{S[table-format=-1.3]}{0.826} & \multicolumn{2}{S[table-format=-1.3]}{0.827} \\
Adj. R\textsuperscript{2} & \multicolumn{2}{S[table-format=-1.3]}{0.805} & \multicolumn{2}{S[table-format=-1.3]}{0.825} & \multicolumn{2}{S[table-format=-1.3]}{0.826} \\
BIC & \multicolumn{2}{S[table-format=-1.3]}{-328.3} & \multicolumn{2}{S[table-format=-1.3]}{-357.6} & \multicolumn{2}{S[table-format=-1.3]}{-354.9} \\
F-statistic & \multicolumn{2}{S[table-format=-1.3]}{1316.9\textsuperscript{***}} & \multicolumn{2}{S[table-format=-1.3]}{750.8\textsuperscript{***}} & \multicolumn{2}{S[table-format=-1.3]}{504.7\textsuperscript{***}} \\
Durbin-Watson & \multicolumn{2}{S[table-format=-1.3]}{2.147} & \multicolumn{2}{S[table-format=-1.3]}{2.168} & \multicolumn{2}{S[table-format=-1.3]}{2.159} \\
Cond. No. & \multicolumn{2}{S[table-format=-1.3]}{121} & \multicolumn{2}{S[table-format=-1.3]}{158} & \multicolumn{2}{S[table-format=-1.3]}{6468} \\
\bottomrule
\end{tabular*}
\hspace*{\fill}
\begin{tablenotes}[center]
\small
\item 
Standard errors in parentheses. \textsuperscript{***}$p < 0.001$, \textsuperscript{**}$p < 0.01$, \textsuperscript{*}$p < 0.1$. Based on 320 observations.
\end{tablenotes}
\end{table}

\begin{table}[h!]
\centering
\caption{Comparison of OLS Regression Models for Theft.}
\label{tab:regression_models_theft}
\hspace*{\fill}
\begin{tabular*}{0.88\textwidth}{l@{\extracolsep{\fill}}*{3}{S[table-format=-1.3]@{\,}l}}
\toprule
 & \multicolumn{2}{c}{{Model 1}} & \multicolumn{2}{c}{{Model 2}} & \multicolumn{2}{c}{{Model 3}} \\
\midrule
Intercept & -3.224\textsuperscript{***} & ($0.197$) & -3.041\textsuperscript{***} & ($0.178$) & 1.876 & ($2.044$) \\
$N$ & 1.221\textsuperscript{***} & ($0.038$) & 0.920\textsuperscript{***} & ($0.048$) & -0.032 & ($0.397$) \\
$I$ &  &  & 0.316\textsuperscript{***} & ($0.035$) & -0.799\textsuperscript{*} & ($0.463$) \\
$N \times I$ &  &  &  &  & 0.215\textsuperscript{*} & ($0.089$) \\
\midrule
R\textsuperscript{2} & \multicolumn{2}{S[table-format=-1.3]}{0.763} & \multicolumn{2}{S[table-format=-1.3]}{0.811} & \multicolumn{2}{S[table-format=-1.3]}{0.815} \\
Adj. R\textsuperscript{2} & \multicolumn{2}{S[table-format=-1.3]}{0.763} & \multicolumn{2}{S[table-format=-1.3]}{0.810} & \multicolumn{2}{S[table-format=-1.3]}{0.813} \\
BIC & \multicolumn{2}{S[table-format=-1.3]}{-265.8} & \multicolumn{2}{S[table-format=-1.3]}{-332.6} & \multicolumn{2}{S[table-format=-1.3]}{-332.7} \\
F-statistic & \multicolumn{2}{S[table-format=-1.3]}{1025.2\textsuperscript{***}} & \multicolumn{2}{S[table-format=-1.3]}{681.4\textsuperscript{***}} & \multicolumn{2}{S[table-format=-1.3]}{463.2\textsuperscript{***}} \\
Durbin-Watson & \multicolumn{2}{S[table-format=-1.3]}{1.874} & \multicolumn{2}{S[table-format=-1.3]}{1.894} & \multicolumn{2}{S[table-format=-1.3]}{1.892} \\
Cond. No. & \multicolumn{2}{S[table-format=-1.3]}{121} & \multicolumn{2}{S[table-format=-1.3]}{158} & \multicolumn{2}{S[table-format=-1.3]}{6468} \\
\bottomrule
\end{tabular*}
\hspace*{\fill}
\begin{tablenotes}[center]
\small
\item 
Standard errors in parentheses. \textsuperscript{***}$p < 0.001$, \textsuperscript{**}$p < 0.01$, \textsuperscript{*}$p < 0.1$. Based on 320 observations.
\end{tablenotes}
\end{table}

\bibliography{sciadvbib}

\begin{thebibliography}{10}
\expandafter\ifx\csname url\endcsname\relax
  \def\url#1{\texttt{#1}}\fi
\expandafter\ifx\csname urlprefix\endcsname\relax\def\urlprefix{URL }\fi
\providecommand{\bibinfo}[2]{#2}
\providecommand{\eprint}[2][]{\url{#2}}

\bibitem{Chamlin2004}
\bibinfo{author}{Chamlin, M.~B.} \& \bibinfo{author}{Cochran, J.~K.}
\newblock \bibinfo{title}{{An Excursus on the Population Size-Crime
  Relationship}}.
\newblock \emph{\bibinfo{journal}{Western Criminology Review}}
  \textbf{\bibinfo{volume}{5}}, \bibinfo{pages}{119--130}
  (\bibinfo{year}{2004}).

\bibitem{Rotolo2006}
\bibinfo{author}{Rotolo, T.} \& \bibinfo{author}{Tittle, C.~R.}
\newblock \bibinfo{title}{{Population Size, Change, and Crime in U.S. Cities}}.
\newblock \emph{\bibinfo{journal}{Journal of Quantitative Criminology}}
  \textbf{\bibinfo{volume}{22}}, \bibinfo{pages}{341--367}
  (\bibinfo{year}{2006}).

\bibitem{oliveira2021more}
\bibinfo{author}{Oliveira, M.}
\newblock \bibinfo{title}{More crime in cities? {On} the scaling laws of crime
  and the inadequacy of per capita rankings—a cross-country study}.
\newblock \emph{\bibinfo{journal}{Crime Science}}
  \textbf{\bibinfo{volume}{10}}, \bibinfo{pages}{1--13} (\bibinfo{year}{2021}).

\bibitem{Gibbs1976}
\bibinfo{author}{Gibbs, J.~P.} \& \bibinfo{author}{Erickson, M.~L.}
\newblock \bibinfo{title}{{Crime Rates of American Cities in an Ecological
  Context}}.
\newblock \emph{\bibinfo{journal}{American Journal of Sociology}}
  \textbf{\bibinfo{volume}{82}}, \bibinfo{pages}{605--620}
  (\bibinfo{year}{1976}).

\bibitem{stafford_crime_1980}
\bibinfo{author}{Stafford, M.~C.} \& \bibinfo{author}{Gibbs, J.~P.}
\newblock \bibinfo{title}{Crime {Rates} in an {Ecological} {Context}:
  {Extension} of a {Proposition}}.
\newblock \emph{\bibinfo{journal}{Social Science Quarterly}}
  \textbf{\bibinfo{volume}{61}}, \bibinfo{pages}{653--665}
  (\bibinfo{year}{1980}).

\bibitem{StultsHasbrouck}
\bibinfo{author}{Stults, M., Brian J.~Hasbrouck}.
\newblock \bibinfo{title}{The effect of commuting on city-level crime rates}.
\newblock \emph{\bibinfo{journal}{Journal of Quantitative Criminology}}
  \textbf{\bibinfo{volume}{31}}, \bibinfo{pages}{331--350}
  (\bibinfo{year}{2015}).

\bibitem{neal_connected_2013}
\bibinfo{author}{Neal, Z.~P.}
\newblock \emph{\bibinfo{title}{The connected city: how networks are shaping
  the modern metropolis}}.
\newblock The metropolis and modern life (\bibinfo{publisher}{Routledge},
  \bibinfo{address}{New York, NY}, \bibinfo{year}{2013}), \bibinfo{edition}{1st
  ed} edn.

\bibitem{liang_intercity_2024}
\bibinfo{author}{Liang, X.}, \bibinfo{author}{Hidalgo, C.~A.},
  \bibinfo{author}{Balland, P.-A.}, \bibinfo{author}{Zheng, S.} \&
  \bibinfo{author}{Wang, J.}
\newblock \bibinfo{title}{Intercity connectivity and urban innovation}.
\newblock \emph{\bibinfo{journal}{Computers, Environment and Urban Systems}}
  \textbf{\bibinfo{volume}{109}}, \bibinfo{pages}{102092}
  (\bibinfo{year}{2024}).

\bibitem{Mayhew1976}
\bibinfo{author}{Mayhew, B.~H.} \& \bibinfo{author}{Levinger, R.~L.}
\newblock \bibinfo{title}{{Size and the Density of Interaction in Human
  Aggregates}}.
\newblock \emph{\bibinfo{journal}{American Journal of Sociology}}
  \textbf{\bibinfo{volume}{82}}, \bibinfo{pages}{86--110}
  (\bibinfo{year}{1976}).

\bibitem{blau1977inequality}
\bibinfo{author}{Blau, P.}
\newblock \emph{\bibinfo{title}{Inequality and Heterogeneity: A Primitive
  Theory of Social Structure}} (\bibinfo{publisher}{Free Press},
  \bibinfo{year}{1977}), \bibinfo{edition}{1} edn.

\bibitem{wirth1938urbanism}
\bibinfo{author}{Wirth, L.}
\newblock \bibinfo{title}{Urbanism as a way of life}.
\newblock \emph{\bibinfo{journal}{American journal of sociology}}
  \textbf{\bibinfo{volume}{44}}, \bibinfo{pages}{1--24} (\bibinfo{year}{1938}).

\bibitem{Groff2015}
\bibinfo{author}{Groff, E.~R.}
\newblock \bibinfo{title}{{Informal Social Control and Crime Events}}.
\newblock \emph{\bibinfo{journal}{Journal of Contemporary Criminal Justice}}
  \textbf{\bibinfo{volume}{31}}, \bibinfo{pages}{90--106}
  (\bibinfo{year}{2015}).

\bibitem{fischer1975toward}
\bibinfo{author}{Fischer, C.~S.}
\newblock \bibinfo{title}{Toward a subcultural theory of urbanism}.
\newblock \emph{\bibinfo{journal}{American Journal of Sociology}}
  \textbf{\bibinfo{volume}{80}}, \bibinfo{pages}{1319--1341}
  (\bibinfo{year}{1975}).

\bibitem{Fischer1995}
\bibinfo{author}{Fischer, C.~S.}
\newblock \bibinfo{title}{{The Subcultural Theory of Urbanism: A Twentieth-Year
  Assessment}}.
\newblock \emph{\bibinfo{journal}{American Journal of Sociology}}
  \textbf{\bibinfo{volume}{101}}, \bibinfo{pages}{543--577}
  (\bibinfo{year}{1995}).

\bibitem{Bettencourt2007}
\bibinfo{author}{Bettencourt, L. M.~A.}, \bibinfo{author}{Lobo, J.},
  \bibinfo{author}{Helbing, D.}, \bibinfo{author}{K{\"u}hnert, C.} \&
  \bibinfo{author}{West, G.~B.}
\newblock \bibinfo{title}{Growth, innovation, scaling, and the pace of life in
  cities}.
\newblock \emph{\bibinfo{journal}{Proceedings of the National Academy of
  Sciences}} \textbf{\bibinfo{volume}{104}}, \bibinfo{pages}{7301--7306}
  (\bibinfo{year}{2007}).

\bibitem{farley_ecological_1981}
\bibinfo{author}{Farley, J.~E.} \& \bibinfo{author}{Hansel, M.}
\newblock \bibinfo{title}{The {Ecological} {Context} of {Urban} {Crime}: {A}
  {Further} {Exploration}}.
\newblock \emph{\bibinfo{journal}{Urban Affairs Quarterly}}
  \textbf{\bibinfo{volume}{17}}, \bibinfo{pages}{37--54}
  (\bibinfo{year}{1981}).

\bibitem{farley_suburbanization_1987}
\bibinfo{author}{Farley, J.~E.}
\newblock \bibinfo{title}{Suburbanization and {Central}-{City} {Crime} {Rates}:
  {New} {Evidence} and a {Reinterpretation}}.
\newblock \emph{\bibinfo{journal}{American Journal of Sociology}}
  \textbf{\bibinfo{volume}{93}}, \bibinfo{pages}{688--700}
  (\bibinfo{year}{1987}).

\bibitem{oliveira2017scaling}
\bibinfo{author}{Oliveira, M.}, \bibinfo{author}{Bastos-Filho, C.} \&
  \bibinfo{author}{Menezes, R.}
\newblock \bibinfo{title}{The scaling of crime concentration in cities}.
\newblock \emph{\bibinfo{journal}{PloS one}} \textbf{\bibinfo{volume}{12}},
  \bibinfo{pages}{e0183110} (\bibinfo{year}{2017}).

\bibitem{hipp_cities_2017}
\bibinfo{author}{Hipp, J.~R.} \& \bibinfo{author}{Kane, K.}
\newblock \bibinfo{title}{Cities and the larger context: {What} explains
  changing levels of crime?}
\newblock \emph{\bibinfo{journal}{Journal of Criminal Justice}}
  \textbf{\bibinfo{volume}{49}}, \bibinfo{pages}{32--44}
  (\bibinfo{year}{2017}).

\bibitem{hipp_accounting_2021}
\bibinfo{author}{Hipp, J.~R.} \& \bibinfo{author}{Williams, S.~A.}
\newblock \bibinfo{title}{Accounting for {Meso}- or {Micro}-{Level} {Effects}
  {When} {Estimating} {Models} {Using} {City}-{Level} {Crime} {Data}:
  {Introducing} a {Novel} {Imputation} {Technique}}.
\newblock \emph{\bibinfo{journal}{Journal of Quantitative Criminology}}
  \textbf{\bibinfo{volume}{37}}, \bibinfo{pages}{915--951}
  (\bibinfo{year}{2021}).

\bibitem{caminha_human_2017}
\bibinfo{author}{Caminha, C.} \emph{et~al.}
\newblock \bibinfo{title}{Human mobility in large cities as a proxy for crime}.
\newblock \emph{\bibinfo{journal}{PLOS ONE}} \textbf{\bibinfo{volume}{12}},
  \bibinfo{pages}{e0171609} (\bibinfo{year}{2017}).

\bibitem{Boggs1965}
\bibinfo{author}{Boggs, S.~L.}
\newblock \bibinfo{title}{{Urban Crime Patterns}}.
\newblock \emph{\bibinfo{journal}{American Sociological Review}}
  \textbf{\bibinfo{volume}{30}}, \bibinfo{pages}{899} (\bibinfo{year}{1965}).

\bibitem{Andresen2006}
\bibinfo{author}{Andresen, M.~A.}
\newblock \bibinfo{title}{{Crime Measures and the Spatial Analysis of Criminal
  Activity}}.
\newblock \emph{\bibinfo{journal}{The British Journal of Criminology}}
  \textbf{\bibinfo{volume}{46}}, \bibinfo{pages}{258--285}
  (\bibinfo{year}{2006}).

\bibitem{Andresen2011}
\bibinfo{author}{Andresen, M.~A.}
\newblock \bibinfo{title}{{The Ambient Population and Crime Analysis}}.
\newblock \emph{\bibinfo{journal}{The Professional Geographer}}
  \textbf{\bibinfo{volume}{63}}, \bibinfo{pages}{193--212}
  (\bibinfo{year}{2011}).

\bibitem{alves_commuting_2021}
\bibinfo{author}{Alves, L. G.~A.}, \bibinfo{author}{Rybski, D.} \&
  \bibinfo{author}{Ribeiro, H.~V.}
\newblock \bibinfo{title}{Commuting network effect on urban wealth scaling}.
\newblock \emph{\bibinfo{journal}{Scientific Reports}}
  \textbf{\bibinfo{volume}{11}}, \bibinfo{pages}{22918} (\bibinfo{year}{2021}).

\bibitem{yang_inter-city_2024}
\bibinfo{author}{Yang, V.~C.}, \bibinfo{author}{Jackson, J.~J.} \&
  \bibinfo{author}{Kempes, C.~P.}
\newblock \bibinfo{title}{Inter-city firm connections and the scaling of urban
  economic indicators}.
\newblock \emph{\bibinfo{journal}{PNAS Nexus}} \textbf{\bibinfo{volume}{3}},
  \bibinfo{pages}{pgae503} (\bibinfo{year}{2024}).

\bibitem{loureiro_impact_2025}
\bibinfo{author}{Loureiro, N.~A.}, \bibinfo{author}{Neto, C.~R.},
  \bibinfo{author}{Sutton, J.}, \bibinfo{author}{Perc, M.} \&
  \bibinfo{author}{Ribeiro, H.~V.}
\newblock \bibinfo{title}{Impact of inter-city interactions on disease
  scaling}.
\newblock \emph{\bibinfo{journal}{Scientific Reports}}
  \textbf{\bibinfo{volume}{15}}, \bibinfo{pages}{498} (\bibinfo{year}{2025}).

\bibitem{louf_how_2014}
\bibinfo{author}{Louf, R.} \& \bibinfo{author}{Barthelemy, M.}
\newblock \bibinfo{title}{How congestion shapes cities: from mobility patterns
  to scaling}.
\newblock \emph{\bibinfo{journal}{Scientific Reports}}
  \textbf{\bibinfo{volume}{4}}, \bibinfo{pages}{5561} (\bibinfo{year}{2014}).

\bibitem{cobb_theory_1928}
\bibinfo{author}{Cobb, C.~W.} \& \bibinfo{author}{Douglas, P.~H.}
\newblock \bibinfo{title}{A {Theory} of {Production}}.
\newblock \emph{\bibinfo{journal}{The American Economic Review}}
  \textbf{\bibinfo{volume}{18}}, \bibinfo{pages}{139--65}
  (\bibinfo{year}{1928}).

\bibitem{varian_intermediate_2014}
\bibinfo{author}{Varian, H.~R.}
\newblock \emph{\bibinfo{title}{Intermediate {Microeconomics} with {Calculus}}}
  (\bibinfo{publisher}{W. W. Norton \& Company}, \bibinfo{address}{New York,
  NY}, \bibinfo{year}{2014}), \bibinfo{edition}{1} edn.

\bibitem{lobo_urban_2013}
\bibinfo{author}{Lobo, J.}, \bibinfo{author}{Bettencourt, L. M.~A.},
  \bibinfo{author}{Strumsky, D.} \& \bibinfo{author}{West, G.~B.}
\newblock \bibinfo{title}{Urban {Scaling} and the {Production} {Function} for
  {Cities}}.
\newblock \emph{\bibinfo{journal}{PLoS ONE}} \textbf{\bibinfo{volume}{8}},
  \bibinfo{pages}{e58407} (\bibinfo{year}{2013}).

\bibitem{ribeiro_effects_2019}
\bibinfo{author}{Ribeiro, H.~V.}, \bibinfo{author}{Rybski, D.} \&
  \bibinfo{author}{Kropp, J.~P.}
\newblock \bibinfo{title}{Effects of changing population or density on urban
  carbon dioxide emissions}.
\newblock \emph{\bibinfo{journal}{Nature Communications}}
  \textbf{\bibinfo{volume}{10}}, \bibinfo{pages}{3204} (\bibinfo{year}{2019}).

\bibitem{ribeiro_mathematical_2023}
\bibinfo{author}{Ribeiro, F.~L.} \& \bibinfo{author}{Rybski, D.}
\newblock \bibinfo{title}{Mathematical models to explain the origin of urban
  scaling laws}.
\newblock \emph{\bibinfo{journal}{Physics Reports}}
  \textbf{\bibinfo{volume}{1012}}, \bibinfo{pages}{1--39}
  (\bibinfo{year}{2023}).

\bibitem{leitao2016scaling}
\bibinfo{author}{Leitao, J.~C.}, \bibinfo{author}{Miotto, J.~M.},
  \bibinfo{author}{Gerlach, M.} \& \bibinfo{author}{Altmann, E.~G.}
\newblock \bibinfo{title}{Is this scaling nonlinear?}
\newblock \emph{\bibinfo{journal}{Royal Society open science}}
  \textbf{\bibinfo{volume}{3}}, \bibinfo{pages}{150649} (\bibinfo{year}{2016}).

\bibitem{altmann_spatial_2020}
\bibinfo{author}{Altmann, E.~G.}
\newblock \bibinfo{title}{Spatial interactions in urban scaling laws}.
\newblock \emph{\bibinfo{journal}{PLOS ONE}} \textbf{\bibinfo{volume}{15}},
  \bibinfo{pages}{e0243390} (\bibinfo{year}{2020}).

\bibitem{Arcaute2014}
\bibinfo{author}{Arcaute, E.} \emph{et~al.}
\newblock \bibinfo{title}{{Constructing cities, deconstructing scaling laws}}.
\newblock \emph{\bibinfo{journal}{Journal of The Royal Society Interface}}
  \textbf{\bibinfo{volume}{12}}, \bibinfo{pages}{20140745--20140745}
  (\bibinfo{year}{2014}).

\bibitem{Cottineau2017}
\bibinfo{author}{Cottineau, C.}, \bibinfo{author}{Hatna, E.},
  \bibinfo{author}{Arcaute, E.} \& \bibinfo{author}{Batty, M.}
\newblock \bibinfo{title}{{Diverse cities or the systematic paradox of Urban
  Scaling Laws}}.
\newblock \emph{\bibinfo{journal}{Computers, Environment and Urban Systems}}
  \textbf{\bibinfo{volume}{63}}, \bibinfo{pages}{80--94}
  (\bibinfo{year}{2017}).

\bibitem{Louf2014}
\bibinfo{author}{Louf, R.} \& \bibinfo{author}{Barthelemy, M.}
\newblock \bibinfo{title}{{Scaling: Lost in the Smog}}.
\newblock \emph{\bibinfo{journal}{Environment and Planning B: Planning and
  Design}} \textbf{\bibinfo{volume}{41}}, \bibinfo{pages}{767--769}
  (\bibinfo{year}{2014}).

\bibitem{white2014spatial}
\bibinfo{author}{White, S.}, \bibinfo{author}{Yehle, T.},
  \bibinfo{author}{Serrano, H.}, \bibinfo{author}{Oliveira, M.} \&
  \bibinfo{author}{Menezes, R.}
\newblock \bibinfo{title}{The spatial structure of crime in urban
  environments}.
\newblock In \emph{\bibinfo{booktitle}{Social Informatics}},
  \bibinfo{pages}{102--111} (\bibinfo{publisher}{Springer},
  \bibinfo{year}{2014}).

\bibitem{10.1371/journal.pone.0143638}
\bibinfo{author}{Davies, T.} \& \bibinfo{author}{Marchione, E.}
\newblock \bibinfo{title}{Event networks and the identification of crime
  pattern motifs}.
\newblock \emph{\bibinfo{journal}{PLOS ONE}} \textbf{\bibinfo{volume}{10}},
  \bibinfo{pages}{1--19} (\bibinfo{year}{2015}).

\bibitem{oliveira2015criminal}
\bibinfo{author}{Oliveira, M.}, \bibinfo{author}{Barbosa-Filho, H.},
  \bibinfo{author}{Yehle, T.}, \bibinfo{author}{White, S.} \&
  \bibinfo{author}{Menezes, R.}
\newblock \bibinfo{title}{From criminal spheres of familiarity to crime
  networks}.
\newblock In \emph{\bibinfo{booktitle}{Complex Networks VI}},
  \bibinfo{pages}{219--230} (\bibinfo{publisher}{Springer},
  \bibinfo{year}{2015}).

\bibitem{Malleson2016}
\bibinfo{author}{Malleson, N.} \& \bibinfo{author}{Andresen, M.~A.}
\newblock \bibinfo{title}{{Exploring the impact of ambient population measures
  on London crime hotspots}}.
\newblock \emph{\bibinfo{journal}{Journal of Criminal Justice}}
  \textbf{\bibinfo{volume}{46}}, \bibinfo{pages}{52--63}
  (\bibinfo{year}{2016}).

\bibitem{PachecoOM17}
\bibinfo{author}{Pacheco, D.~F.}, \bibinfo{author}{Oliveira, M.} \&
  \bibinfo{author}{Menezes, R.}
\newblock \bibinfo{title}{Using social media to assess neighborhood social
  disorganization: {A} case study in the {U}nited {K}ingdom}.
\newblock In \emph{\bibinfo{booktitle}{Proceedings of the Thirtieth
  International Florida Artificial Intelligence Research Society Conference,
  {FLAIRS} 2017, Marco Island, Florida, USA, May 22-24, 2017.}},
  \bibinfo{pages}{341--346} (\bibinfo{year}{2017}).

\bibitem{Commutedataset}
\bibinfo{author}{{Office for National Statistics}}.
\newblock \bibinfo{title}{2011 {Place} of {Residence} by {Place} of {Work},
  {Local} {Authority}} (\bibinfo{year}{2014}).
\newblock
  \urlprefix\url{https://data.london.gov.uk/dataset/place-residence-place-work-local-authority}.
\newblock \bibinfo{note}{Accessed: 2025-05-12}.

\bibitem{Crimedataset}
\bibinfo{author}{{Home Office}}.
\newblock \bibinfo{title}{Police recorded crime {Community} {Safety}
  {Partnership} {Open} {Data} {Table}, from year ending {March} 2012 to year
  ending {March} 2015} (\bibinfo{year}{2015}).
\newblock
  \urlprefix\url{https://www.gov.uk/government/statistics/police-recorded-crime-open-data-tables}.
\newblock \bibinfo{note}{Accessed: 2025-05-12}.

\bibitem{LADtoCSPdataset}
\bibinfo{author}{{Office for National Statistics}}.
\newblock \bibinfo{title}{{LAD} to {Community Safety Partnership} to {PFA}
  ({December} 2019) {Lookup} in {EW}} (\bibinfo{year}{2019}).
\newblock
  \urlprefix\url{https://geoportal.statistics.gov.uk/datasets/8bd726c1cd5340e3980d03e0877efb3e_0}.
\newblock \bibinfo{note}{Accessed: 2025-05-12}.

\bibitem{LADInformationDataset}
\bibinfo{author}{mySociety}.
\newblock \bibinfo{title}{Uk local authorities (past and current)}
  (\bibinfo{year}{2024}).
\newblock
  \urlprefix\url{https://pages.mysociety.org/uk_local_authority_names_and_codes/downloads/uk-la-past-current-uk-local-authorities-current-csv/latest#survey}.
\newblock \bibinfo{note}{Accessed: 2025-05-26}.

\bibitem{PopulationDataset}
\bibinfo{author}{{Office for National Statistics}}.
\newblock \bibinfo{title}{2011 {Census}: {Key} {Statistics} for {Local}
  {Authorities} in {England} and {Wales}} (\bibinfo{year}{2012}).
\newblock
  \urlprefix\url{https://www.ons.gov.uk/peoplepopulationandcommunity/populationandmigration/populationestimates/datasets/2011censuskeystatisticsforlocalauthoritiesinenglandandwales}.
\newblock \bibinfo{note}{Accessed: 2025-05-12}.

\end{thebibliography}
\bibliographystyle{naturemag}
 
\noindent 
\section*{Declarations}

\subsection*{Availability of data and materials}
All data used in this study are publicly available through official sources, as detailed in the Methods section.

\subsection*{Competing interests}
The authors declare no competing interests.

\subsection*{Funding}
This research received no external funding.




\end{document}